\documentclass[sigconf, screen]{acmart}
\acmConference[ISSTA 2024]{ACM SIGSOFT International Symposium on Software Testing and Analysis}{16-20 September, 2024}{Vienna, Austria}

\AtBeginDocument{%
  }

\setcopyright{acmcopyright}
\copyrightyear{2024}
\acmYear{2024}
\acmDOI{XXXXXXX.XXXXXXX}

\acmPrice{}
\acmISBN{978-1-4503-XXXX-X/18/06}

\usepackage{amsmath}
\usepackage{tikz,eso-pic}
\usepackage{amsfonts}
\usepackage{xspace} 
\usepackage{mathrsfs}
\usepackage{amsmath}
\usepackage{amsthm}
\usepackage{subfigure}
\usepackage{booktabs}
\usepackage{multirow}
\usepackage{graphicx}
\usepackage{flushend}
\usepackage{pifont}
\usepackage{caption}
\usepackage{float}
\usepackage[normalem]{ulem}
\usepackage{enumitem}

\usepackage{bbding}
\usepackage{makecell}
\usepackage{listings}
\usepackage{threeparttable}

\usepackage[linesnumbered,ruled]{algorithm2e}
\RestyleAlgo{ruled}
\SetKwComment{Comment}{* }{ */}
\SetKwInOut{Input}{Input}
\SetKwInOut{Output}{Output}

\SetKwFunction{EmitLaser}{EmitLaser}
\SetKwFunction{PointGeneration}{PointGeneration}
\SetKwFunction{GetReflectivity}{GetReflectivity}
\SetKwFunction{Attenuation}{Attenuation}
\SetKwFunction{Reflection}{Reflection}
\SetKwFunction{InitUniqueScenario}{InitUniqueScenario}
\SetKwFunction{Simulation}{Simulation}
\SetKwFunction{RandomMutate}{RandomMutate}
\SetKwFunction{InitTemperature}{InitTemperature}
\SetKwFunction{ReportCornerCase}{ReportCornerCase}
\SetKwFunction{Append}{Append}
\SetKwFunction{Random}{Random}
\SetKwFunction{IsCornerCase}{IsCornerCase}

\newcommand{\sys}{\textsc{AutoSVT}\xspace}

\begin{document}

\title{First-principles Based 3D Virtual Simulation Testing for Discovering SOTIF Corner Cases of Autonomous Driving}


\author{Lehang Li}
\affiliation{
	\institution{Harbin Institute of Technology, Shenzhen}
	\city{Shenzhen}
	\country{China}
}
\email{lilehang1@gmail.com}

\author{Haokuan Wu}
\affiliation{
	\institution{Harbin Institute of Technology, Shenzhen}
	\city{Shenzhen}
	\country{China}
}
\email{thisiskkya@gmail.com}

\author{Botao Yao}
\affiliation{
	\institution{Harbin Institute of Technology, Shenzhen}
	\city{Shenzhen}
	\country{China}
}
\email{yaobt-hit@outlook.com}

\author{Tianyu He}
\affiliation{
	\institution{Harbin Institute of Technology, Shenzhen}
	\city{Shenzhen}
	\country{China}
}
\email{tianyuhe2017@gmail.com}

\author{Shuohan Huang}
\affiliation{
	\institution{Harbin Institute of Technology, Shenzhen}
	\city{Shenzhen}
	\country{China}
}
\email{huangsh_2022@163.com}

\author{Chuanyi Liu}
\affiliation{
	\institution{Harbin Institute of Technology, Shenzhen}
	\city{Shenzhen}
	\country{China}
}
\email{liuchuanyi@hit.edu.cn}

\renewcommand{\shortauthors}{Li et al.}

\begin{abstract}
    3D virtual simulation, which generates diversified test scenarios and tests full-stack of Autonomous Driving Systems (ADSes) modules dynamically as a whole, is a promising approach for Safety of The Intended Functionality (SOTIF) ADS testing. However, as different configurations of a test scenario will affect the sensor perceptions and environment interaction, e.g. light pulses emitted by the LiDAR sensor will undergo backscattering and attenuation, which is usually overlooked by existing works, leading to false positives or wrong results. Moreover, the input space of an ADS is extremely large, with infinite number of possible initial scenarios and mutations, along both temporal and spatial domains. 
  
    This paper proposes a first-principles based sensor modeling and environment interaction scheme, and integrates it into CARLA simulator. With this scheme, a long-overlooked category of adverse weather related corner cases are discovered, along with their root causes. Moreover, a meta-heuristic algorithm is designed based on several empirical insights, which guide both seed scenarios and mutations, significantly reducing the search dimensions of scenarios and enhancing the efficiency of corner case identification. Experimental results show that under identical simulation setups, our algorithm discovers about four times as many corner cases as compared to state-of-the-art work.
	
\end{abstract}




\maketitle

\section{Introduction}
\label{sec:introduction}

Autonomous vehicles should never go to public until they have high confidence on Safety of The Intended Functionality (SOTIF)~\cite{iso21448}. 
Safety assurance concerns whether autonomous vehicles may encounter safety risks in various scenarios, implying a shift from mileage focused testing to scenario-coverage focused testing. 
Scenarios with safety risks often follow a long-tail distribution, making it challenging to encounter such situations in the real world. Taking adverse weather conditions for example, either waiting for a heavy rainstorm or artificially creating a dense fog environment brings substantial cost.
As an alternative to testing in the real world, 3D virtual simulation is a promising approach for generating diversified test scenarios and testing full-stack of Autonomous Driving System (ADS) modules dynamically as a whole. However, as different configurations of a test scenario will affect the sensor perceptions and interaction with the environment, e.g. light pulses emitted by the LiDAR sensor will undergo backscattering and attenuation, which is usually overlooked by existing works, leading to false negative or wrong results. Moreover, the input space of an ADS is extremely large, with infinite number of possible initial scenarios and mutations, along both temporal and spatial domains. 

There are a bunch of related works on discovering accident scenarios, i.e. corner cases, in 3D virtual simulators ~\cite{fremont2020formal,scanlon2021waymo,leaderboard, drivefuzz,hu2021coverage,avfuzzer,autofuzz}, which are categorized and analyzed in section 2.2. However, there remains several technical challenges which are summarized as follows:  

\textbf{C1. High-fidelity sensor modeling and interaction with the environment.} Existing 3D virtual simulators, such as CARLA~\cite{dosovitskiy2017carla} and LGSVL~\cite{rong2020lgsvl}, simulate weather primarily for visual effects, neglecting the influence of weather on LiDAR. Adverse weather conditions, such as fog, lead to backscattering and attenuation in laser signals emitted by LiDAR, resulting in a reduction in the accuracy of object detection algorithms~\cite{rasshofer2011influences, fog_simulation,weather_keisuke2019,linnhoff2022measuring,zang2019impact}. Taking foggy conditions as an example, whether addressing accuracy through retraining LiDAR-based algorithms with data augmentation techniques using adverse weather data~\cite{fog_simulation,snowfall_simulation} or adopting the Multi-Sensor Fusion (MSF) approach to compensate for the limitations of a single sensor~\cite{msf_wu2022,msf_yang2022}, the accuracy of LiDAR-based object detection algorithms remains notably low under adverse weather conditions. Therefore, there is a need to conduct safety assessments of ADS in foggy weather. 

Current research predominantly emphasizes the perception module~\cite{hadj2019lidar,koschmieder1924theorie,yang2020lanoising,yang2021performance,lee2022gan,teufel2022simulating,fog_simulation,bijelic2020seeing,rasshofer2011influences,snowfall_simulation}, relying on static datasets for testing. However, this approach is limited to simulating point cloud datasets, overlooking the interaction between LiDAR and the environment. Furthermore, it lacks scenario diversity and can only assess the perception module of an ADS.

However, ADSes encompass more than just the perception module. To assess the safety of ADSes in adverse weather, testing solely on the perception module using datasets is inadequate. Errors in the perception module can either be amplified or mitigated by other modules. For instance, if the perception module outputs a bounding box for a vehicle 30 meters away that is much smaller than reality, it may not necessarily lead to erroneous planning by the planning module. Additionaly, testing focused only on the perception module uses non-interactive data, where the route of the ego vehicle remains fixed regardless of the output of perception module. However, in real-world scenarios, the perception output at time $t$ influences the input of planning and control, which subsequently affects the input of perception at time $t+1$. Consequently, interactive testing has the potential to magnify perception errors. 

\textbf{C2. Efficient corner case discovery methods in high dimensional scenario spaces.}  Autonomous driving scenarios encompass a myriad of variable parameters, creating a high dimensional search space that defies complete exploration. Approaches like DriveFuzz~\cite{drivefuzz} integrate traditional fuzzing with corner case discovery, using the initial scenario as a seed with mutations affecting scenario parameters. While traditional fuzzing relies on code coverage as feedback to guide mutation and seed selection, SOTIF testing isn't about exhaustively covering all code in an ADS; rather, it aims to generate diverse scenarios. Consequently, fuzzing-like methods are unsuitable for effective SOTIF testing.

Despite these approaches redefining code coverage to apply algorithms in autonomous driving scenarios, their efficiency in discovering corner cases remains low. Furthermore, these methods identify corner cases with little interrelation, providing limited assistance in improving ADSes.

\textbf{C3. Complexity of characteristics and module communications of the simulator or ADS.} Specific factors are neglected in combining the simulator and ADS, especially the synchronization mechanisms between ADS and simulator. A synchronization mechanism is needed to ensure that the clock frequency of the ADS is the same as the simulator in co-simulation. High-fidelity rendering increases computational overhead, causing the simulator to be unable to maintain real-time simulation which results in:
\begin{equation}
	\label{eq:time_inconsistent}
	r = \frac{T_{\text{sim}}}{T_{\text{wall}}} < 1
\end{equation}

Here, $T_{\text{sim}}$ and $T_{\text{wall}}$ denote the durations in the simulation and wall clock, respectively. In this case, the working frequency of modules in the ADS referencing the wallclock will be much higher than the simulation frequency of the simulator. This leads to a discrepancy between the performance of the ADS in virtual simulation and its real-world performance. As errors accumulate over time, it might eventually result in false positives or false negatives~\cite{hu2021disclosing}.

\textbf{Contributions}.
To address these challenges, this paper makes the following contributions:
\begin{itemize}
    \item A first-principles based sensor modeling and interaction with the environment is proposed, which is further integrated into CARLA simulator. In our simulation, the long-overlooked safety issue in autonomous driving related to adverse weather is exposed.
    \item Different with existing fuzzing based methods, a meta-heuristic algorithm is designed based on several empirical insights. These insights guide both seed scenarios and mutations, significantly reducing the search dimensions of scenarios and enhancing the efficiency of corner case identification. Experimental results demonstrate that, under identical simulation environments and configurations, our algorithm discovers about four times as many corner cases in 100 simulation tests compared to the current state-of-the-art work (DriveFuzz). All discovered corner cases and related videos are showcased at \url{https://idslab-autosec.github.io/}.
    \item An eBPF-based synchronization mechanism is designed to bridge ADS and the simulator. This addresses the discrepancies in clock frequencies between the simulator and ADSes, significantly mitigating issues of false positives and negatives arising from asynchronous simulations in the domain.
    \item The simulation testing platform, named \sys, encompasses first-principles based simulation, the corner case discovery algorithm, and the synchronization mechanism. It is open-sourced at \url{https://github.com/idslab-autosec/AutoSVT}.
\end{itemize}

\section{Related Work}
\label{sec:related_work}

\subsection{Sensor Modeling and Interaction with the Environment}
The primary goal of sensor modeling is to achieve a high-fidelity representation of real-world environments. In existing 3D virtual simulators, fog simulation is presently confined to image-based methods, affecting only camera data and not influencing LiDAR point clouds. However, previous research~\cite{linnhoff2022measuring,weather_keisuke2019,zang2019impact} demonstrates the significant impact of adverse weather conditions on LiDAR. 

Yang et al.~\cite{yang2020lanoising,yang2021performance} utilize fog chamber data to train neural networks for LiDAR sensor performance, enabling the simulation of LiDAR signal strength in various foggy conditions. Lee et al.~\cite{lee2022gan} propose using a Generative Adversarial Network (GAN) to learn LiDAR data mapping relationships under different weather conditions, predicting point cloud distributions in foggy conditions. However, data-driven methods heavily depend on the quality and scale of the dataset, and collecting data for adverse weather conditions is challenging. Moreover, probability-based models often lack physical interpretability and may significantly deviate from actual outcomes due to real-world data generalization.

In addition to data-driven approaches, some research utilizes physics-based modeling methods for simulations. Hadj-Bachir et al. and Teufel et al.~\cite{teufel2022simulating,hadj2019lidar} propose physics-based models for the impact of foggy conditions on LiDAR raycasting. Nevertheless, because these models rely on empirical rules and probabilistic approximations of physical formulas, their fidelity is limited.

Furthermore, the first-principles based simulation approach is rooted in fundamental laws and physical principles. Consequently, this method can provide highly realistic sensor simulation data at the particle level. As the state-of-the-art research on LiDAR simulation in foggy weather, Hahner et al.~\cite{fog_simulation} analyse the entire process from LiDAR signal transmission to reception. This study introduces a method for generating highly realistic LiDAR data under varying fog densities, using clear weather LiDAR data. 

However, a limitation of~\cite{fog_simulation} is its reliance on post-processing static LiDAR data from datasets. This approach cannot accurately capture the real-time dynamic information of LiDAR interacting with the environment, such as the reflection and scattering of individual laser beams on different materials. Moreover, while~\cite{fog_simulation} primarily focuses on evaluating the impact of fog on LiDAR object detection algorithms, it is insufficient to just assess these perception algorithms in isolation. A comprehensive evaluation of ADSes requires a holistic approach, taking into account all modules and components. 

\subsection{Corner Case Discovery}

%

Research on corner case discovery in simulators can be mainly categorized into two types. The first type involves replaying real-world high-risk driving scenarios or accident data in a simulator~\cite{fremont2020formal,scanlon2021waymo,leaderboard}. This approach can only validate the performance of an ADS in known safety-critical scenarios within the simulator but cannot discover new corner cases.


The second approach generates new test scenarios in the simulator that do not exist in reality. Existing works primarily adopt a fuzzing approach~\cite{drivefuzz,hu2021coverage,avfuzzer,autofuzz}. In this method, scenarios are treated as seeds for fuzzing, and the mutation of the seed corresponds to the mutation of scenario parameters. Traditional fuzzing methods are typically guided by code coverage, but in the context of autonomous driving simulation, code coverage is not applicable.
As a pioneering study, AV-Fuzzer~\cite{avfuzzer} uses the lateral and longitudinal safety distances of the ego vehicle as feedback to guide scenario mutation. 
However, AV-Fuzzer only considers simple changes in the behavior of vehicles near the ego vehicle, neglecting the mutation of scenario elements such as maps and weather.
ASF~\cite{hu2021coverage} employs trajectory coverage as feedback, aiming to have the ego vehicle's trajectory cover as many blocks in the map as possible. However, trajectory coverage only reflects the relationship between the ego vehicle and the map, lacking a comprehensive evaluation of the interaction between the ego vehicle and traffic flow.
Drivefuzz~\cite{drivefuzz} and AutoFuzz~\cite{autofuzz} represent state-of-the-art works. Drivefuzz uses metrics related to the comfort of driving, such as ego vehicle acceleration and steering angle, and the minimum distance between the ego vehicle and other vehicles as weighted feedback, discovering 30 corner cases. AutoFuzz~\cite{autofuzz} employs a neural network for seed scenario selection and scenario evaluation, using gradients to guide scenario mutation. 
Despite the advancements, the efficiency of discovering corner cases in fuzzing-based methods remains suboptimal. Moreover, these methods often discover unrelated corner cases, making it challenging to analyze the root causes of these cases and providing limited assistance in improving ADSes.
\section{Design}
\label{sec:design}

\subsection{Overview}
\label{sec:overview}
\begin{figure}[h]
	\centerline{\includegraphics[width=0.5\textwidth]{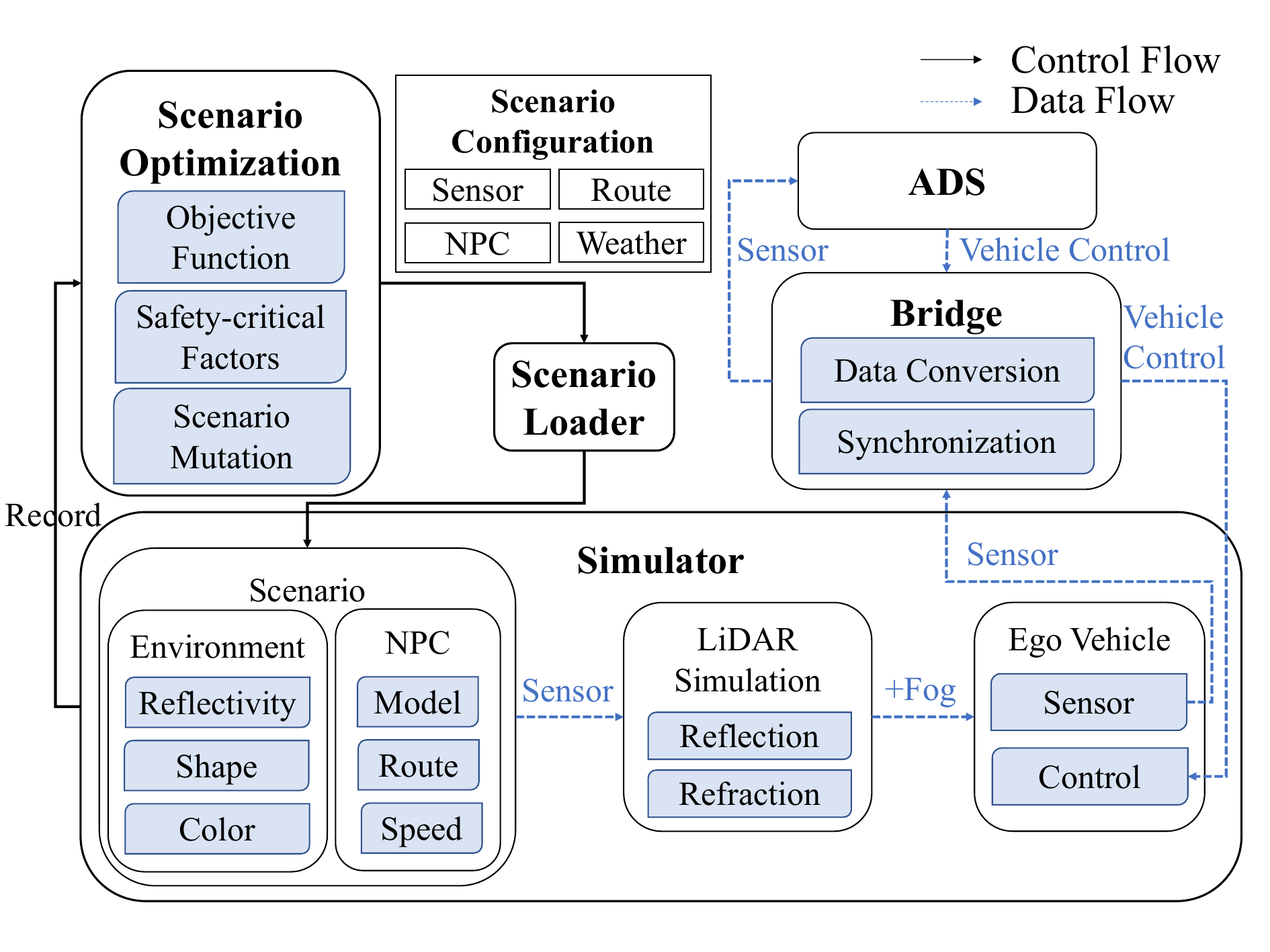}}
	\caption{Overview of \sys.}
	\label{fig:overview}
\end{figure}

\sys enables SOTIF testing of all ADS modules under adverse weather conditions and can discover a multitude of corner cases, particularly in foggy weather. The workflow of \sys is illustrated in \autoref{fig:overview}.

First, the \textbf{scenario optimization} module (\autoref{sec:scenario_generation}) generates the initial scenario configuration. The scenario configuration comprises four main elements: the sensor configuration of the ego vehicle, the route of the ego vehicle (i.e., the starting and destination points), Non-Player Characters (NPCs) in the scenario (including vehicles and pedestrians), and the weather conditions. The scenario configuration is then output to the \textbf{scenario loader}, which calls the simulator's API to generate a 3D virtual scenario.

The \textbf{simulator} then performs simulations. During a simulation, the environment transmits the reflectivity, shape, and color of objects in the scenario to the sensors of the ego vehicle. At this stage, our first-principles based LiDAR simulation algorithm (\autoref{sec:fog_simulation}) calculates the reflection and refraction during the LiDAR signal transmission process, and then transmits the point cloud data in adverse weather to the ego vehicle. 

The ego vehicle, through the \textbf{bridge}, transfers sensor data to the ADS. The bridge serves as the component connecting the ADS and simulator, responsible for data conversion and synchronization (\autoref{sec:synchronization}). Upon receiving sensor data, the ADS ultimately generates control commands for the ego vehicle in the simulator. 

After each simulation run, the data collected during the test is transferred to the scenario optimization module, which calculates the objective function value for that simulation and optimizes the scenario based on several safety-critical factors to generate scenarios more prone to corner cases.

\subsection{First-principles Based 3D Virtual Simulation}
\label{sec:fog_simulation}

\begin{figure}[htbp]
	\centerline{\includegraphics[width=0.5\textwidth]{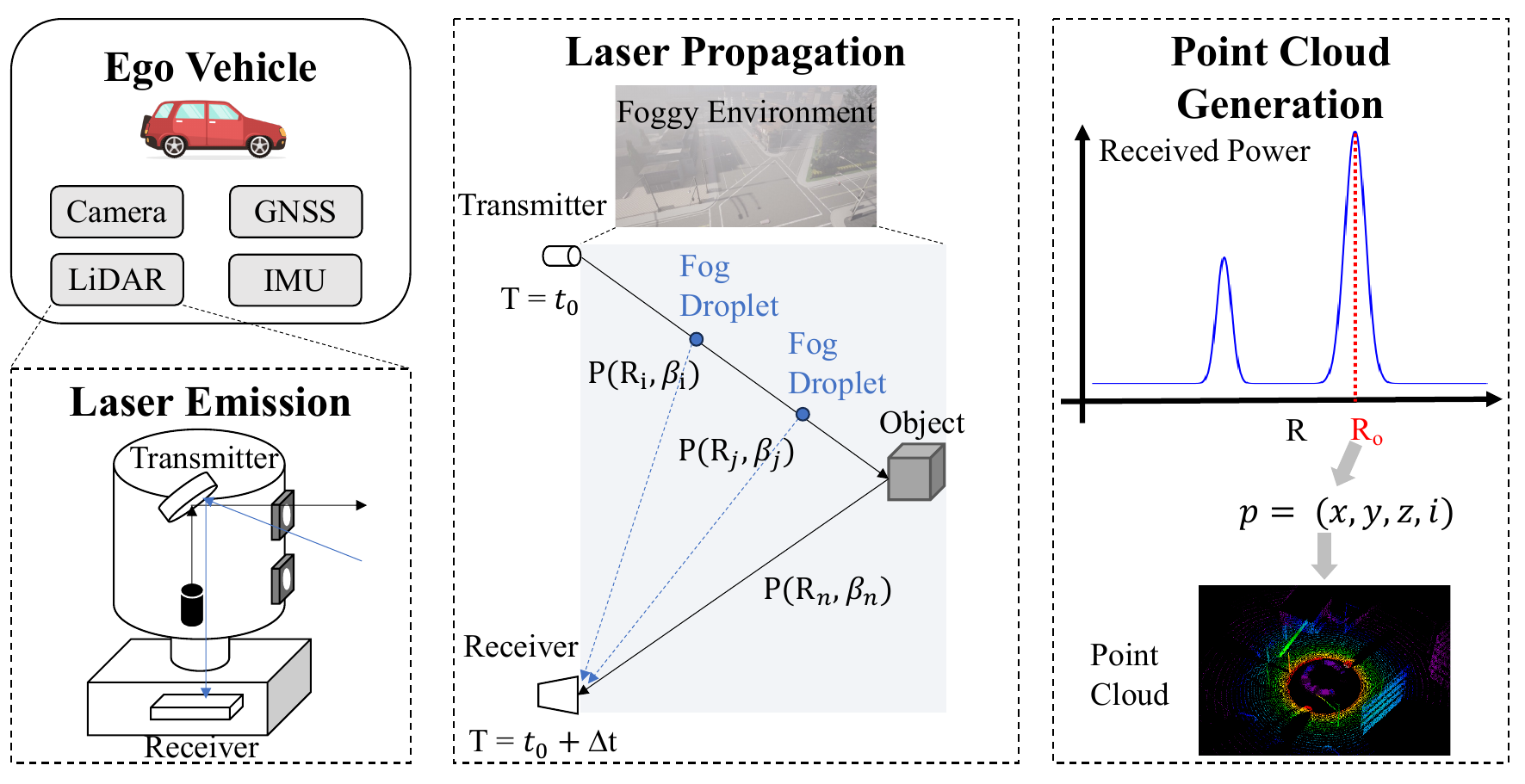}}
	\caption{The procedure of first-principles based LiDAR simulation in foggy weather.}
	\label{fig:sensor_simulation}
\end{figure}

We propose a first-principles based LiDAR simulation algorithm capable of accurately simulating LiDAR data in foggy weather. The simulation algorithm can be divided into three stages: laser emission, laser propagation, and point cloud generation, as illustrated in \autoref{fig:sensor_simulation}.

\textbf{Laser Emission.} In our simulation, LiDAR rotates 360 degrees and emits lasers. The direction of each laser is determined by the LiDAR's horizontal and vertical angular resolutions. When the LiDAR rotates to a specific horizontal angle, it simultaneously emits multiple lasers to cover the vertical field of view with vertical angular resolution.

\textbf{Laser Propagation.} A laser in the environment may undergo multiple reflections, but except for the first reflection, other reflections have a minimal impact on the point cloud~\cite{hanke2017generation}. Therefore, in our simulation, when a laser hits the first hard target, we no longer consider subsequent reflections.

As shown in \autoref{fig:sensor_simulation}, when the laser encounters fog droplets, a portion of the energy is reflected back, while another portion is attenuated and penetrates the fog droplets, continuing to propagate forward. When the laser hits a hard target, it undergoes diffuse reflection, and during the process of reflecting back to the LiDAR receiver, it further undergoes attenuation.

Rasshofer et al.~\cite{rasshofer2011influences} provide the one-way transmission loss (\autoref{eq:one_way_loss}) during laser propagation in the real world, describing attenuation during the propagation process.
\begin{equation}
	\label{eq:one_way_loss}
	T(R) = \exp\left(-\int_{r=0}^{R}\alpha(r)dr\right)
\end{equation}
In \autoref{eq:one_way_loss}, $\alpha(r)$ represents the attenuation coefficient at distance $r$, and $R$ represents the maximum distance of one-way laser propagation. However, in the simulated world, space is not entirely continuous; voxels represent the smallest elements, so \autoref{eq:one_way_loss} should be rewritten as:
\begin{equation}
	\label{eq:one_way_loss_sim}
	T(R) = \exp\left(-\sum\alpha(r)\Delta r\right) = \exp\left(-\Delta r \sum \alpha(r)\right)
\end{equation}
where $\Delta r$ is the side length of the voxel. The signal of the transmit pulse for an automotive LiDAR can be modeled using \autoref{eq:lidar_pulse}~\cite{rasshofer2011influences}.
\begin{equation}
	\label{eq:lidar_pulse}
	P_T(t)=
	\begin{cases}
		P_0\sin^2(\dfrac{\pi}{2\tau_H}), 0\leq t\leq2\tau_H \\
		0, \text{otherwise}
	\end{cases} 
\end{equation}
Here, $P_0$ denotes the peak power of the LiDAR pulse, and $\tau_H$ represents the half-power pulse width. Lasers are transmitted at light speed $c$, navigating through the simulation world before their reflection. As described by \autoref{eq:one_way_loss_sim}, the laser experiences an attenuation given by $\exp(2\Delta r \sum \alpha(r))$ after its round trip. Our model postulates that the laser's initial encounter is with a target object $o$ situated at a distance $R_0$. The reflective coefficient, $\beta_0(o)$, of this object can be derived from the simulation environment. The signal intensity, $P_{\text{object}}$, from the object's reflection is computed as:
\begin{small}
	\begin{equation}
		\label{eq:hard}
		P_{R,\text{fog}}^{\text{hard}}(R)=
		\begin{cases}
			\begin{aligned}
				
				\dfrac{C_A P_0 \exp(-2\Delta r \sum \alpha(r))}{R_0^2} \times&\\
				
				\beta_0(o)\sin^2\left( \dfrac{\pi (R-R_0)}{c\tau_H}\right)&,R_0 \leq R \leq R_0+c\tau_H\\
				
				0&,\text{otherwise}
				
			\end{aligned}
		\end{cases}
	\end{equation}
\end{small}
Here, $C_A$ denotes a system-specific constant, while $\beta($o$)$ signifies the reflectivity of the target object $o$.

Apart from causing attenuation to the laser's signal strength, atmospheric fog introduces additional reflections. Consequently, the LiDAR-captured signal integrates both the reflections from the target object and the scattering by fog droplets. The intensity of signals reflected due to fog is articulated as:
\begin{small}
	\begin{equation}
		\label{eq:soft}
		\begin{aligned}
			P_{R,\text{fog}}^{\text{soft}}(R)=&C_A P_0 \int_0^{2\tau_H}\sin^2\left(\dfrac{\pi}{2\tau_H}t\right)\times \\
			&\dfrac{\exp(-2\alpha(R-\frac{ct}{2})(R-\frac{ct}{2}))}{(R-\frac{ct}{2})^2} \beta(R-\frac{ct}{2}) \times\\
			&\xi(R-\frac{ct}{2}) U(R_0-R+\frac{ct}{2}) dt
		\end{aligned}
	\end{equation}
\end{small}
Herein, $\xi(R)$ defines the crossover function and $\beta(r)$ encapsulates the backscattering coefficient at a given distance $r$. Ultimately, the aggregate signal intensity registered by the LiDAR is expressed as:
\begin{equation}
	\label{eq:received_signal}
	P_{R,\text{fog}}(R) = P_{R,\text{fog}}^{\text{hard}}(R) + P_{R,\text{fog}}^{\text{soft}}(R)
\end{equation}
The complete calculation methodology can be found in ~\cite{fog_simulation}.

\textbf{Point Cloud Generation.} \autoref{eq:received_signal} is a curve representing the signal intensity received by the LiDAR from a distance $R$. As shown in \autoref{fig:sensor_simulation}, in foggy conditions, the LiDAR receives signals with two peaks, one originating from the target object and the other from fog droplets. When generating a point cloud, the distance corresponding to the maximum signal intensity is selected as the distance to the target object, as indicated in \autoref{eq:r_selection} and \autoref{eq:i_selection}.

\begin{equation}
	\label{eq:r_selection}
	R_o = \mathop{\arg\max}\limits_{R}P_{R,\text{fog}}(R)
\end{equation}
\begin{equation}
	\label{eq:i_selection}
	i = P_{R,\text{fog}}(R_o)
\end{equation}

Here, $R_o$ represents the final measured distance to the target object by the LiDAR, and $i$ represents the intensity of the points. Therefore, when the signal intensity from fog droplets exceeds that of the target object, the fog noise points replace the points belonging to the target object.

\subsection{Synchronization}
\label{sec:synchronization}
We design and implement a RunTime Infrastructure (RTI) compliant with the High Level Architecture (HLA)~\cite{fujimoto1998time} standard to serve as middleware. This middleware, called BPFRTI, enables the collaborative simulation of both the ADS and the simulator. By leveraging eBPF, we adeptly manage the temporal aspects of ADSes that use the system clock as their reference. This solution eliminates the need for complex adaptations or alterations to existing ADS code, offering native support for time synchronization across a wide range of heterogeneous systems.

\begin{figure}[h]
	\centerline{\includegraphics[width=0.36\textwidth]{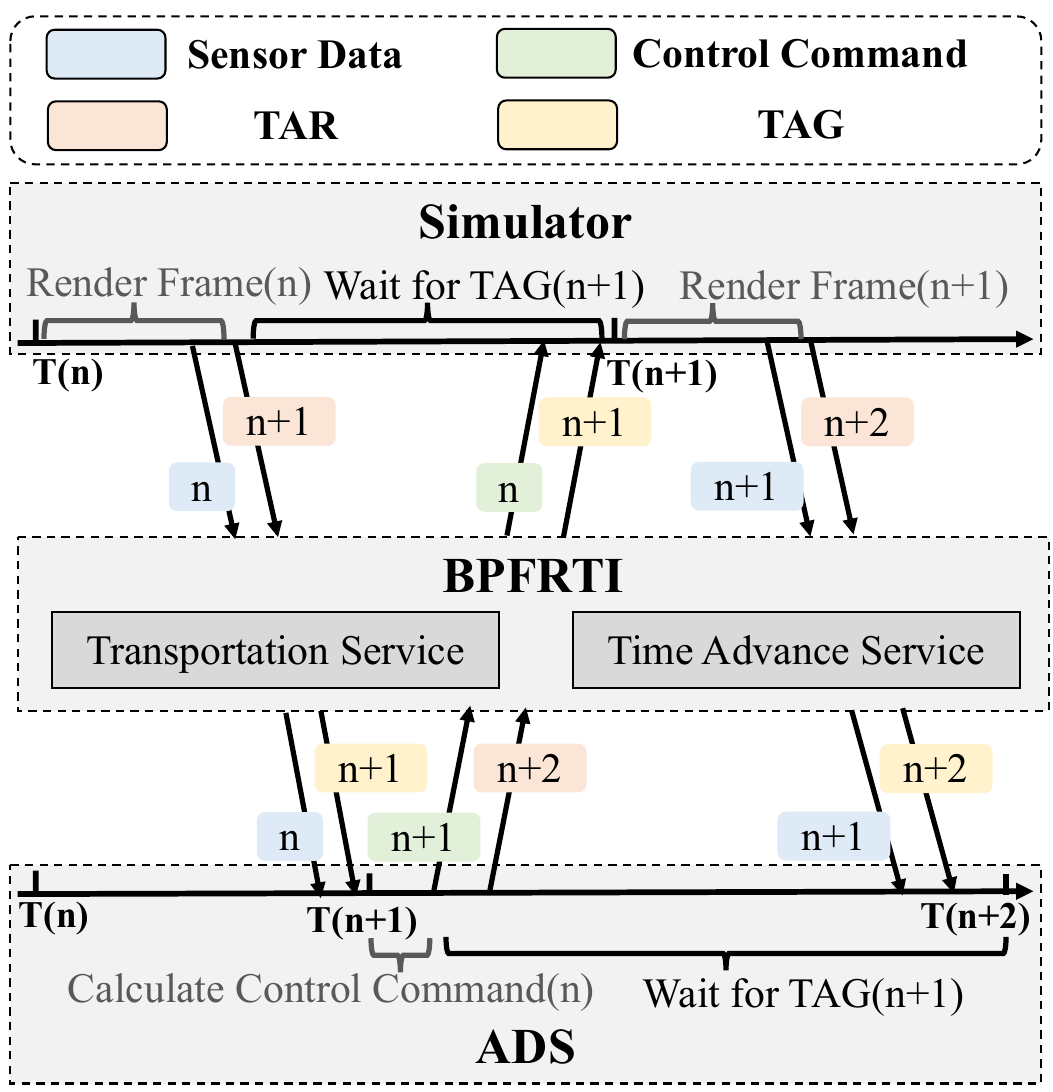}}
	\caption{Workflow of BPFRTI.}
	\label{fig:bpfrti}
\end{figure}

The BPFRTI comprises two components: the transportation service and the time advance service, as illustrated in \autoref{fig:bpfrti}.

\textbf{Transportation Service.} Ensuring the correct sequential transmission of data between ADS and the simulator, we control the order of message transmission based on message timestamps. After rendering a frame of data, the simulator sends the ego vehicle's sensor data to the BPFRTI. Subsequently, the BPFRTI decides whether to transmit this data based on its timestamp. Upon receiving sensor data, the ADS invokes modules such as perception, localization, planning, and control, ultimately generating a control command. This command is initially sent to the BPFRTI, which awaits the simulator's time to advance before forwarding it to the simulator to control the ego vehicle.

\textbf{Time Advance Service.} We employ the time advance protocol to synchronize the time between the simulator and ADS, ensuring synchronization. However, industrial-grade ADSes, such as Apollo, fetch the current system time through system calls to maintain real-time operations. To seamlessly accommodate diverse ADSes without delving into the intricacies of their internal modules, we leverage eBPF technology to capture system calls at the kernel level. Each invocation of such a call triggers the eBPF program, ensuring that the system call returns the simulator's time, thereby achieving synchronization between ADS and simulator times.

The time advance protocol comprises two types of messages: the Time Advance Request (TAR) and the Time Advance Grant (TAG).
\begin{enumerate}
	\item TAR: After the simulator completes rendering a frame of data or ADS finishes computing a control command, it sends data to the BPFRTI. Subsequently, it sends a TAR, requesting time advancement by a fixed time step.
	\item TAG: When the BPFRTI dispatches control commands from ADS to the simulator or forwards sensor data from the simulator to ADS, it sends a TAG, signifying the granting of time advancement. Upon receiving TAG, the simulator advances time by a fixed time step, initiating the rendering of the next frame.
\end{enumerate}

\autoref{fig:bpfrti} presents the complete workflow of the BPFRTI. Upon rendering the sensor data for the n-th frame, the simulator transmits it to the BPFRTI and concurrently dispatches TAR(n+1), requesting a time advancement. Subsequently, the BPFRTI forwards the sensor data to the ADS and sends TAG(n+1), permitting the ADS to advance its time to $(n+1)\Delta t$, where $\Delta t$ denotes the fixed time step. After the ADS computes the control command, it dispatches TAR(n+2) to the BPFRTI and awaits TAG(n+2). The BPFRTI updates the control commands for the ego vehicle in the simulator and sends TAG(n+1), allowing the simulator to progress in time to $(n+1)\Delta t$. Following this, the simulator initiates the computation of the next frame's sensor data.

In \autoref{eq:time_inconsistent}, we employ $\Delta t_{\text{sim}}$ to denote the simulation time step and $\Delta t_{\text{wall}, i}$ to represent the real-time needed for computing the ith frame. With the integration of BPFRTI, we replace the ADS time with eBPF, achieving $\Delta t_{\text{sim}} = \Delta t_{\text{ADS}}$. Although the real-time computations for each frame in the simulator still vary, for the ADS, $r$ remains constant at 1, as demonstrated in \autoref{eq:time_sync}.
\begin{equation}\label{eq:time_sync}
	r = \frac{T_{\text{sim}}}{T_{\text{ADS}}} = \frac{n \Delta t_{\text{sim}}} {n \Delta t_{\text{ADS}}} = 1
\end{equation}

\subsection{Discovering Corner Cases}
\label{sec:scenario_generation}
We have observed that adverse weather conditions such as fog have a significant impact on ADSes. We conduct a series of experiments in a fog simulation environment based on first principles and find that under specific conditions, the accident rate of autonomous vehicles in fog increases significantly. Specifically, we have identified five safety-critical factors that contribute to the increased accident rate in foggy conditions.
\begin{enumerate}
	\item \textbf{Factor1}: The speed of the NPC vehicle in front of the ego vehicle is lower than that of the ego vehicle. When the ego vehicle detects the NPC at a greater distance, it begins to slow down. However, as they get closer, the distance between them continues to decrease. When the NPC enters the noisy area of fog surrounding the ego vehicle, the detection accuracy significantly decreases. This may lead to the ego vehicle not detecting the NPC, subsequently accelerating, and resulting in a collision.
	\item \textbf{Factor2}: The reflection of standing water on the road surface can lead to false positives in the ego vehicle's detection system. This means that the perception module detects obstacles in front of the vehicle that do not actually exist, preventing the ego vehicle from moving forward. This is more likely to occur at night and in foggy conditions.
	\item \textbf{Factor3}: Foggy weather conditions negatively impact the LiDAR's accuracy in detecting small targets, including pedestrians, bicycles, motorcycles, and smaller vehicles. This can result in collisions.
	\item \textbf{Factor4}: Foggy weather also reduces the accuracy of LiDAR in detecting taller vehicles such as firetrucks and SUVs. This is because the noise generated by fog typically falls within the range of 4.5 to 4.7 meters above the ground, which coincides with the height of taller vehicles, further affecting the accuracy of target detection and recognition.
	\item \textbf{Factor5}: In uphill and downhill scenarios, the distribution of fog noise points changes significantly with the terrain, leading to a further reduction in the ego vehicle's perception target detection accuracy. According to the formula in \autoref{sec:fog_simulation} of this paper, when the ego vehicle is moving uphill or downhill, fog noise points form a semi-circular pattern on the uphill side, as shown in  \autoref{fig:slope_pc}.
\end{enumerate}
We design a corner case discovery algorithm based on these five factors, utilizing simulated annealing~\cite{SA} to generalize scenarios.

\subsubsection{Scenarios Initialization}
\label{sec:init_scenario}
Each scenario primarily includes the map, weather conditions, starting and ending points of the ego vehicle, and the type, speed, starting, and ending points of each NPC. Each NPC in the scenario is controlled by the CARLA traffic manager, capable of basic autonomous driving while yielding to other vehicles and adhering to traffic rules. We select several constraints from the aforementioned five factors when initializing scenarios. For example, when selecting factor 3, we limit the generated NPC types to pedestrians, bicycles, and small vehicles. 

To enhance scenario diversity, we generate initial scenarios that are different from previous ones. Zhong et al.~\cite{autofuzz} proposed a definition of a "unique scenario" based on the similarity of scenario description files. This definition avoids generating highly similar scenarios, such as adding an irrelevant pedestrian to a corner case. However, this definition cannot accurately determine whether a scenario is of the same type as the original scenario when multiple unrelated pedestrians are added to a corner case, as there are many different parameters in the scenario.

We use the starting and ending points of the ego vehicle to straightforwardly determine whether two scenarios are of the same type. Specifically, if the starting and ending points of the ego vehicle are different, we consider the two scenarios to be distinct. Our insight is that when the occurrence locations of two corner cases are different, they are evidently dissimilar scenarios.

\subsubsection{Scenarios Mutation}
\label{sec:SA}
We employ the simulated annealing algorithm~\cite{SA} to optimize scenarios, as detailed in \autoref{alg:simulated_annealing}. The algorithm takes a series of scenario features as input, corresponding to the factors in \autoref{sec:scenario_generation}. The output consists of multiple unique corner cases.

\begin{algorithm}
	\caption{Simalated annealing algorithm for corner case discovery}
	\label{alg:simulated_annealing}
	\SetKwInput{KwInput}{Input}
	\SetKwInput{KwOutput}{Output}
	\DontPrintSemicolon
	\Input{Several safety-critical factors $F = \{f_1, f_2, \dots\}$}
	\Output{A series of unique corner cases $S = \{s_1, s_2, \dots\}$}
	$S = \{\}$ \; 
	\For{$i$ $\leftarrow 0$  \KwTo \text{MAX\_SCENARIO}}{
		$s$ = \InitUniqueScenario{$S, F$} \;
		$T$ = \InitTemperature{} \;
		\While{$T \ge$ \text{T\_MIN} \textnormal{\textbf{and}} \textnormal{not} \IsCornerCase{$s$}}{
			$o$ = \Simulation($s$) \;
			\For{$j$ $\leftarrow 0$  \KwTo \text{MAX\_CYCLE}}{
				$s'$ = \RandomMutate{$s, T, F$} \;
				$o'$ = \Simulation($s'$) \;
				\If{\IsCornerCase{$s'$}}{
					$s = s'$ \;
					\ReportCornerCase{$s$} \;
					\Append{$S, s$} \;
					\textbf{break} \;
				}
				\If{$o' \textless o$ \textnormal{\textbf{or}} \Random{0,1}$< \dfrac{e^{o-o'}}{T}$}{
					$s = s'$ \;
				}
			}
			$T = r \times T$ \;
		}
	}
	$\textbf{return}$ $S$
\end{algorithm}

In each iteration, we first initialize a unique scenario using the method described in \autoref{sec:init_scenario} (line 3). Next, we conduct a simulation test on the initial scenario (line 6). The result of the simulation test is the objective function value $o$, which will be explained in \autoref{sec:objective_function}. Within the loop at line 7, the scenario is mutated (line 8), guided by the current temperature $T$ and the set of safety-critical factors $F$. This mutation can take one of four forms:
\begin{itemize}
	\item Adding a new NPC.
	\item Altering the model of an NPC in the scenario. For instance, changing a motorcycle to a truck when the NPC is a vehicle, or modifying the model of a pedestrian.
	\item Adjusting the speed of an NPC in the scenario.
	\item Modifying weather parameters, including sun altitude angles and road surface water. To uncover more fog-related corner cases, the fog concentration for all scenarios is set at 80\% (corresponding to a visibility of 30 meters).
\end{itemize}
The algorithm probabilistically determines whether to mutate the scenario based on $F$ using a temperature-related probability. If the mutation is not based on $F$, the scenario undergoes a random alteration. If based on $F$, for instance, selecting factor3, the model of a specific NPC in the scenario might change to a smaller target. 

The probability of adding a new NPC is higher when $T$ is larger, and the magnitude of scenario mutation is greater. Subsequently, we conduct another simulation test on the mutated scenario (line 9) and compare the objective function values before and after mutation. If the mutated scenario's objective function value is lower, it indicates that the mutated scenario is closer to a corner case. In this case, the mutation is accepted outright. Otherwise, the mutation is accepted with a probability of $\exp{(\dfrac{o’-o}{T})}$ (lines 16-18). A higher temperature $T$ increases the likelihood of accepting mutation with objective function values lower than the original scenario. Finally, we reduce the temperature $T$ using a cooling factor $r\in (0,1)$ (line 20).

\subsubsection{Objective Function}
\label{sec:objective_function}
In the scenario optimization algorithm, the ultimate goal is to find corner cases. However, we cannot use the occurrence of a corner case directly as the algorithm's objective function. Instead, we need more granular metrics. We consider the detection accuracy of the perception module, the physical effects of fog on LiDAR, and the ego vehicle's performance during the driving process, defining the objective function as follows:

\begin{equation}
	\label{eq:objective_function}
	O(s)=-(c_1 \frac{n_{\text{fp}}+n_{\text{fn}}+n_{\text{fog}}}{n_{\text{frame}}} + c_2 \frac{1}{d_{\text{min}}})
\end{equation}

Here, $s$ represents a scenario, $n_{\text{fp}}$ indicates the number of false positives in target detection by the ego vehicle during the simulation, $n_{\text{fn}}$ represents the number of false negatives, $n_{\text{fog}}$ denotes the number of frames in which NPCs are within the fog noise point range around the ego vehicle, $n_{\text{frame}}$ is the total number of simulation frames, $d_{\text{min}}$ signifies the minimum distance between NPCs and the ego vehicle during the simulation, and $c_1$ and $c_2$ are two positive constants.

The first term in \autoref{eq:objective_function} describes the impact of fog on the ego vehicle's perception module during the simulation. The second term reflects the collision probability of the ego vehicle. A decrease in the objective function value indicates that the ego vehicle's perception detects more errors or that the distance between the ego vehicle and NPCs is shorter, indicating a higher likelihood of collision.
\section{Evaluation}
\label{sec:evaluation}

\subsection{Experimental Setup}

We run simulations on a server configured with Ubuntu 20.04, which is powered by dual Intel(R) Xeon(R) Silver 4210R CPUs, 256GB of primary memory, and 8 NVIDIA GeForce RTX 3080 graphics cards. The version of CARLA is 0.9.14. We select the following ADSes as test targets.

\begin{itemize}
	\item We choose Apollo OpenSource 8.0 as one of our test target. Apollo is an open-source industrial-grade ADS that integrates modules such as perception, planning, control. It has received widespread recognition in both industry and academia.
	\item We select four end-to-end autonomous driving algorithms (InterFuser~\cite{shao2023interfuser}, TransFuser~\cite{transfuser}, and TF++~\cite{tf++}) from the CARLA Leaderboard~\cite{leaderboard} as our test targets. We will introduce the CARLA Leaderboard in \autoref{sec:leaderboard}.
\end{itemize}

\subsection{Synchronization}
\label{sec:sync_exp}
In this section, we examine BPFRTI's performance when faced with simulations lagging behind real-time. The larger the gap between simulation time and wallclock time, the more pronounced the desynchronization effects become. To accentuate these discrepancies, we change the simulation's sensor settings to intensify the computational demands for each frame and decreased the timestep accordingly. 

\begin{figure}[htbp]
	\centering
	\subfigure[Without synchronization]{
		\includegraphics[width=0.42\textwidth]{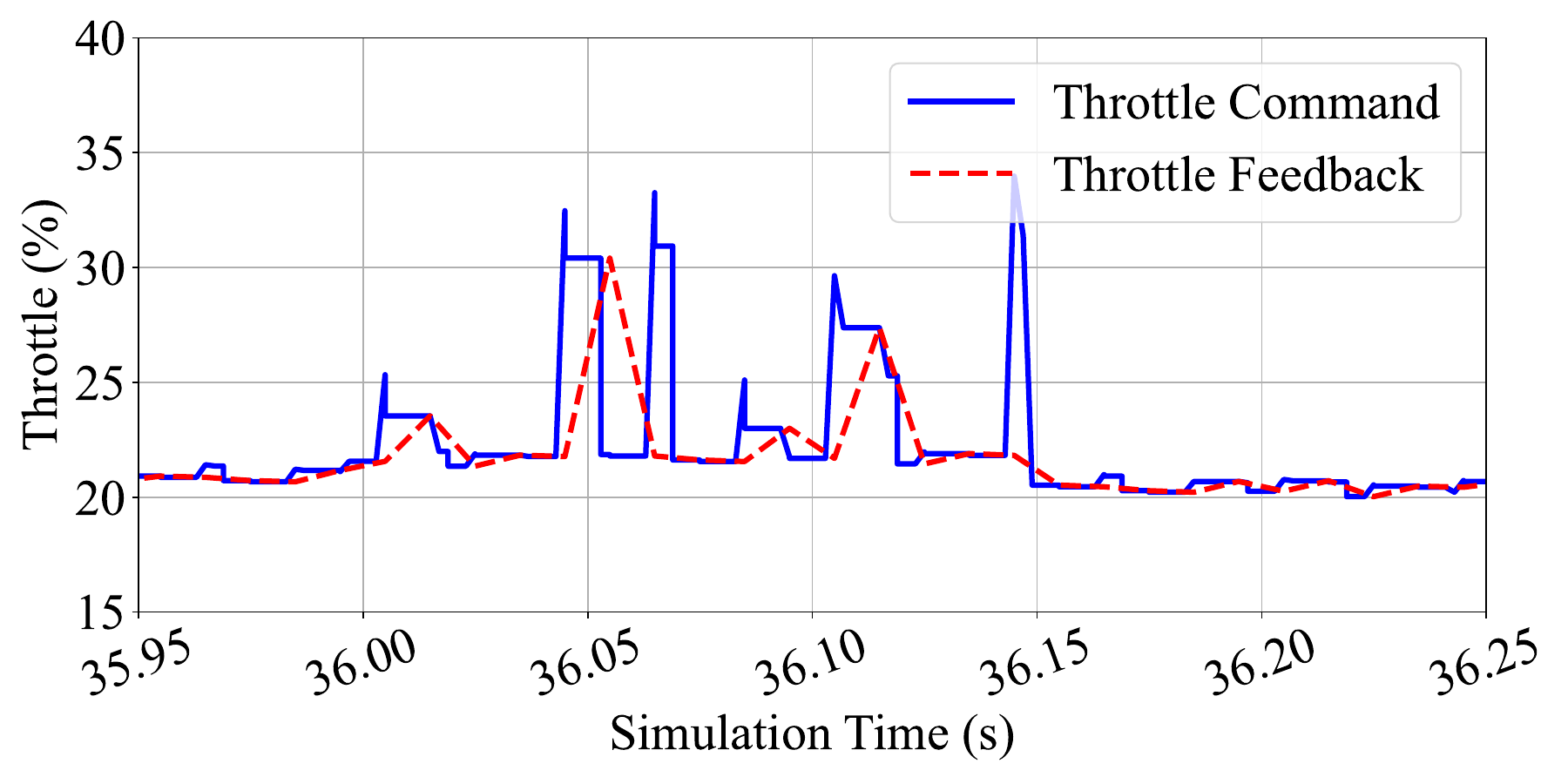}
		\label{fig:async_feedback}
	}
	\subfigure[With synchronization]{
		\includegraphics[width=0.42\textwidth]{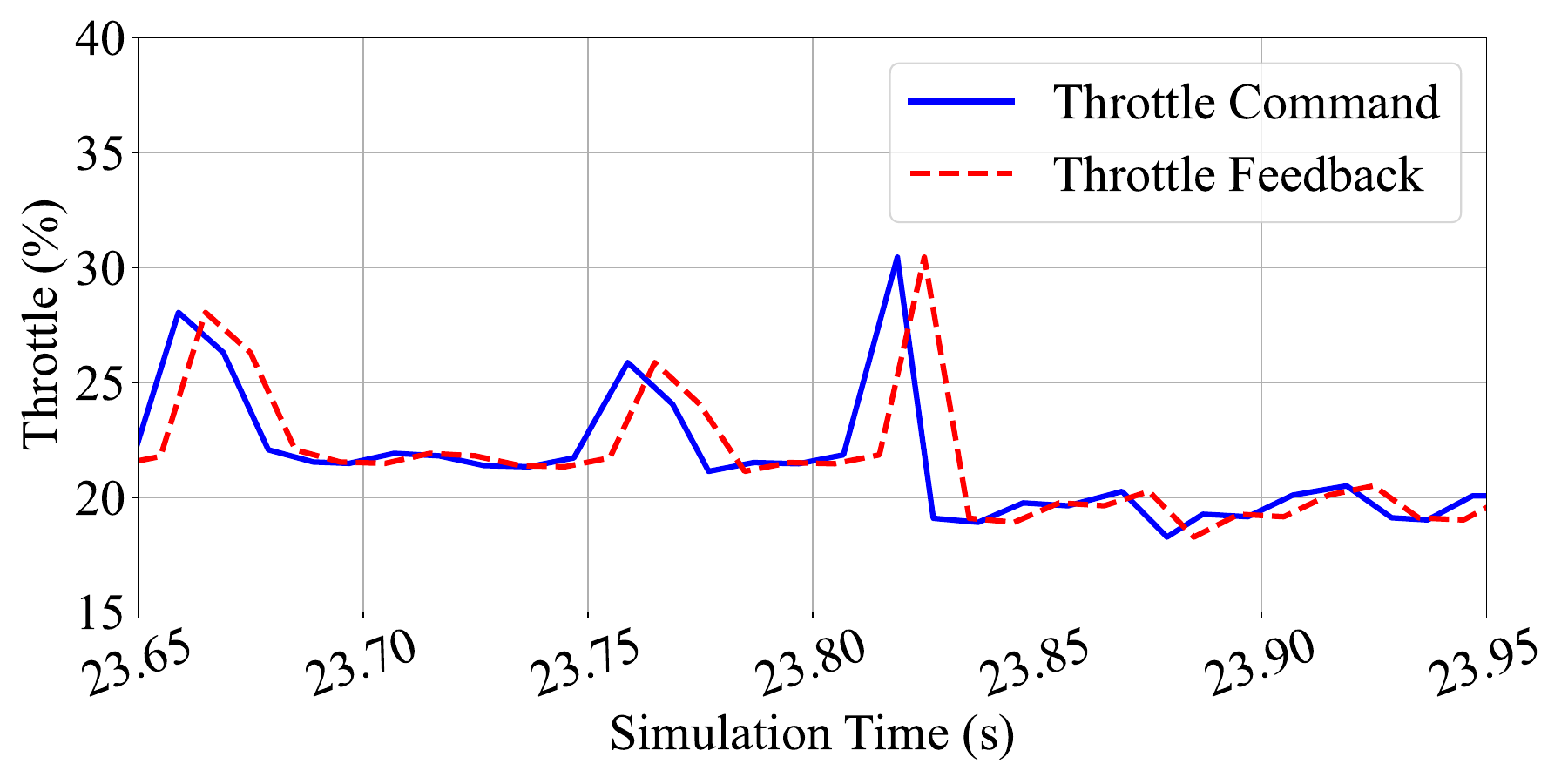}
		\label{fig:sync_feedback}
	}
	
	\caption{Throttle control commands and feedback in Apollo with and without synchronization.}
	\label{fig:sync_exp}
\end{figure}

As depicted in \autoref{fig:async_feedback}, without synchronization, the control module relies on wallclock time as its reference. This results in the generation of throttle commands at an anomalously high rate compared to simulation time. The throttle feedback reveals that numerous commands fail to activate, leading to a misalignment between vehicle chassis data and the issued control commands. On the other hand, \autoref{fig:sync_feedback} illustrates that with the BPFRTI-based time synchronization, there's an almost perfect alignment between throttle commands and feedback, with just a one-time-step latency. This delay occurs because a command initiated in one tick is only reflected in canbus feedback in the subsequent tick.

\subsection{Point Clouds in Fog}
\label{sec:exp_point_cloud}

We integrate the algorithm from \autoref{sec:fog_simulation} into CARLA and investigate the impact of fog on LiDAR point cloud data and object detection algorithms. The experimental results reveal that as fog density increases, the point cloud range decreases, point cloud intensity attenuates more rapidly with distance, and the accuracy of object detection algorithms decreases.

 We use meteorological optical range (MOR) to represent fog density. Following the setup in Hahner et al.~\cite{fog_simulation}, we set $\textnormal{MOR} = \ln{20}/\alpha$, where $\alpha$ represents the fog attenuation coefficient. We also set the backscattering coefficient $\beta = 0.046/\textnormal{MOR}$ and the half-power pulse width $\tau_H$ to be 20 ns.

\begin{figure}[htbp]
	\centering
	\subfigure[Clear weather]{
		\includegraphics[width=0.2\textwidth]{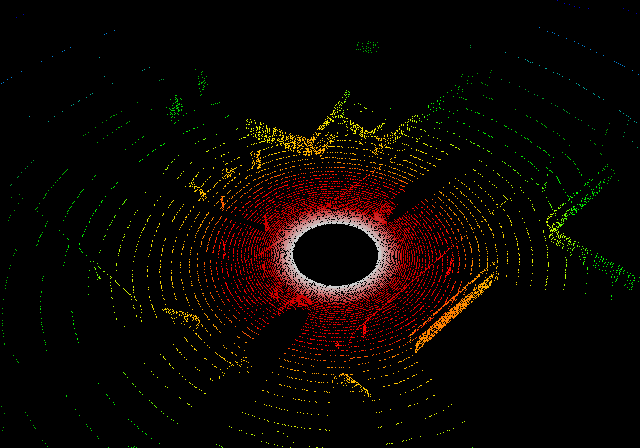}
		\label{fig:clear_pc}
	}
	\subfigure[200m visibility]{
		\includegraphics[width=0.2\textwidth]{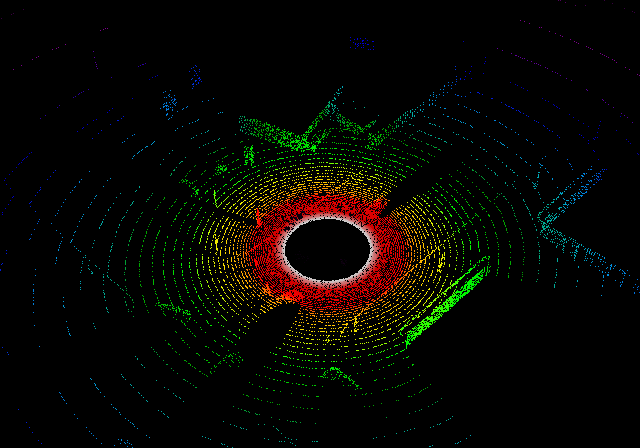}
		\label{fig:200m_pc}
	}
	
	\subfigure[50m visibility]{
		\includegraphics[width=0.2\textwidth]{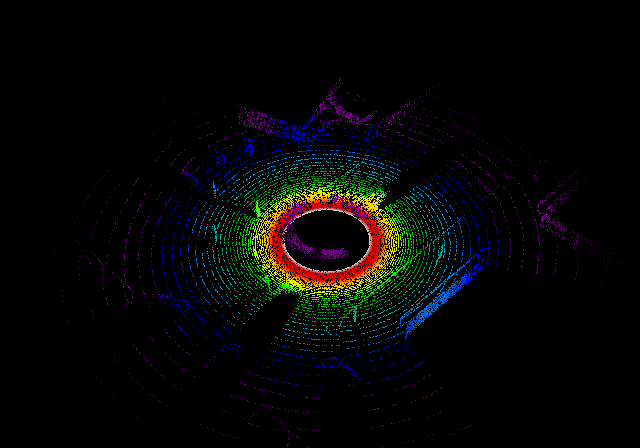}
		\label{fig:50m_pc}
	}
	\subfigure[20m visibility]{
		\includegraphics[width=0.2\textwidth]{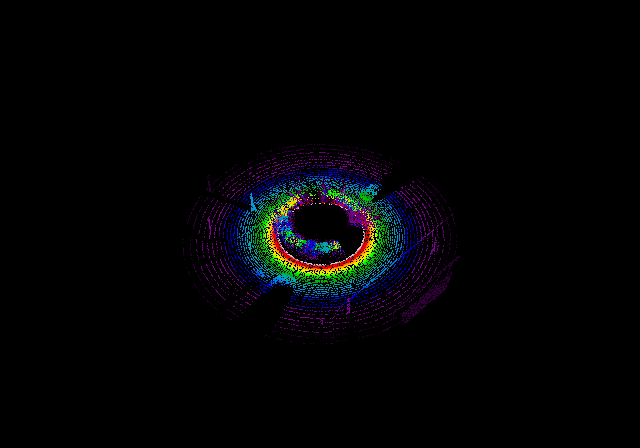}
		\label{fig:20m_pc}
	}
	
	\caption{Point clouds in junction under various fog densities.}
	\label{fig:point_cloud_foggy}
\end{figure}

\autoref{fig:point_cloud_foggy} shows the variation in point clouds in different fog densities. The color indicates point cloud intensity, with higher intensities closer to red and lower intensities closer to purple. On a clear day, the point cloud covers most of the junction (\autoref{fig:clear_pc}). As fog density increases, the range of point cloud rapidly diminishes, ultimately shrinking to a small circle around the ego vehicle. Based on the fog generation patterns discussed in \autoref{sec:fog_simulation}, when fog density is high, the intensity of fog reflections may surpass the intensity of signals returned by distant targets. In such cases, fog noise points may replace genuine target points. Consequently, the direction around the LiDAR with more open space is more prone to generating fog noise points. In \autoref{fig:50m_pc}, we can observe fog noise points appearing around the vehicle when visibility is 50 meters, and the intensity of fog noise points increases further when visibility is reduced to 20 meters (\autoref{fig:20m_pc}).

\begin{figure}[]
	\centering
	\subfigure[]{
		\includegraphics[width=0.2\textwidth]{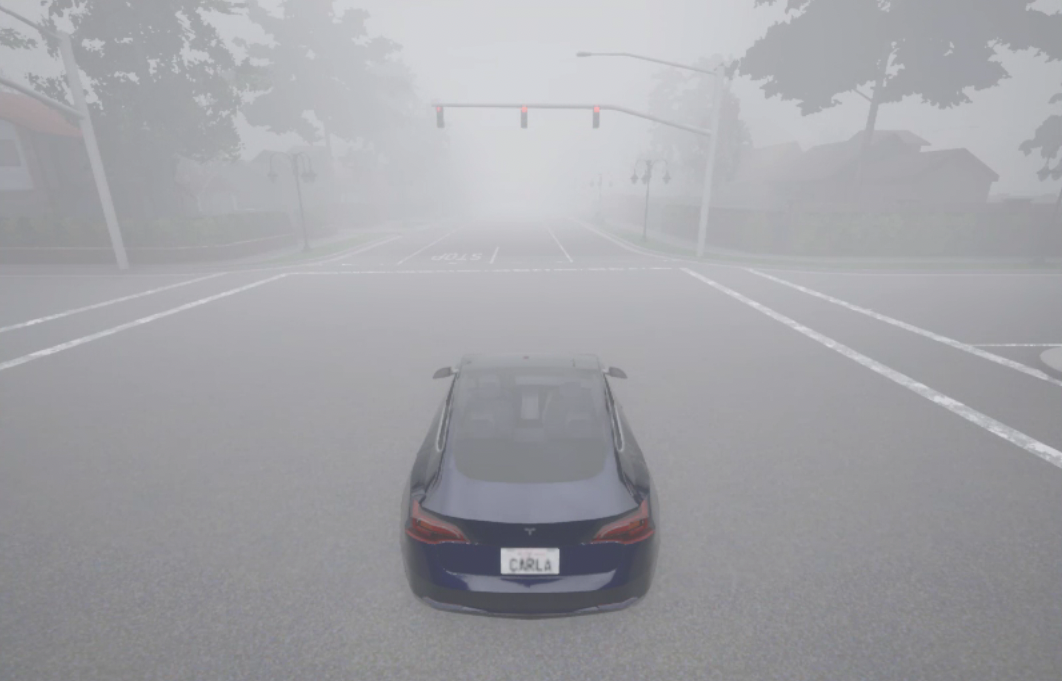}
		\label{fig:junction_pc}
	}
	\subfigure[]{
		\includegraphics[width=0.2\textwidth]{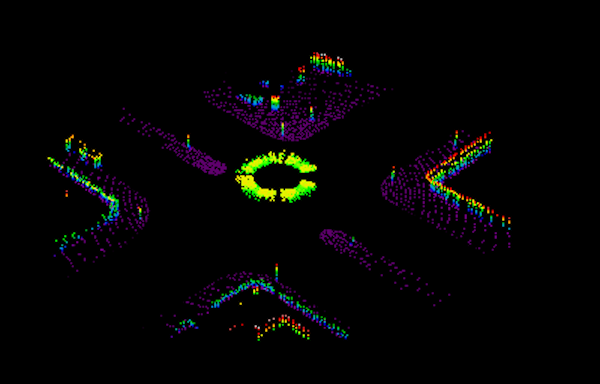}
		\label{fig:junction_camera}
	}
	
	\subfigure[]{
		\includegraphics[width=0.2\textwidth]{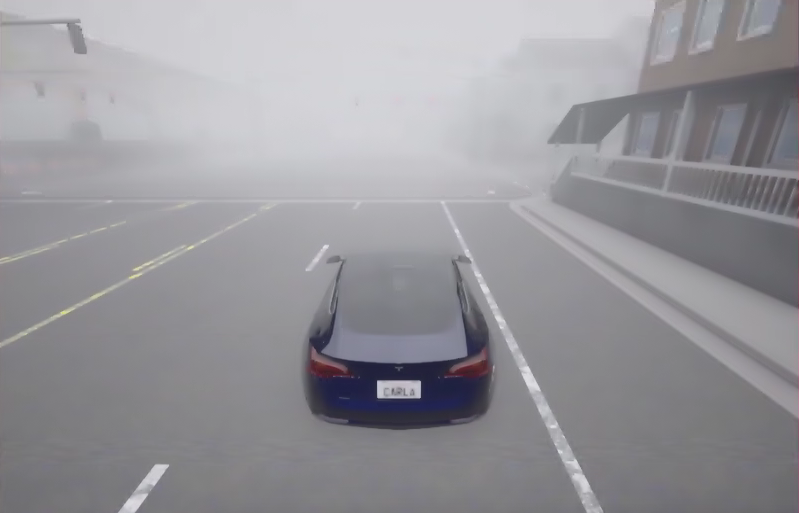}
		\label{fig:bridge_pc}
	}
	\subfigure[]{
		\includegraphics[width=0.2\textwidth]{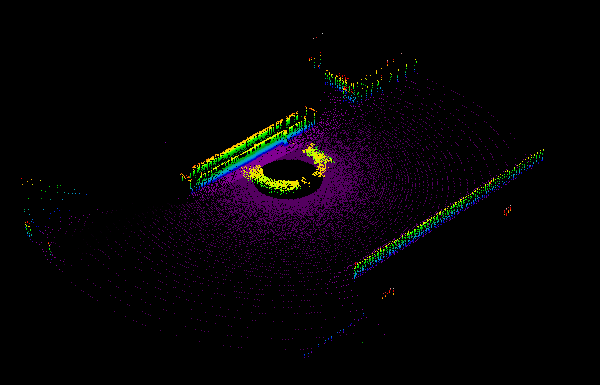}
		\label{fig:bridge_camera}
	}
	
	\subfigure[]{
		\includegraphics[width=0.2\textwidth]{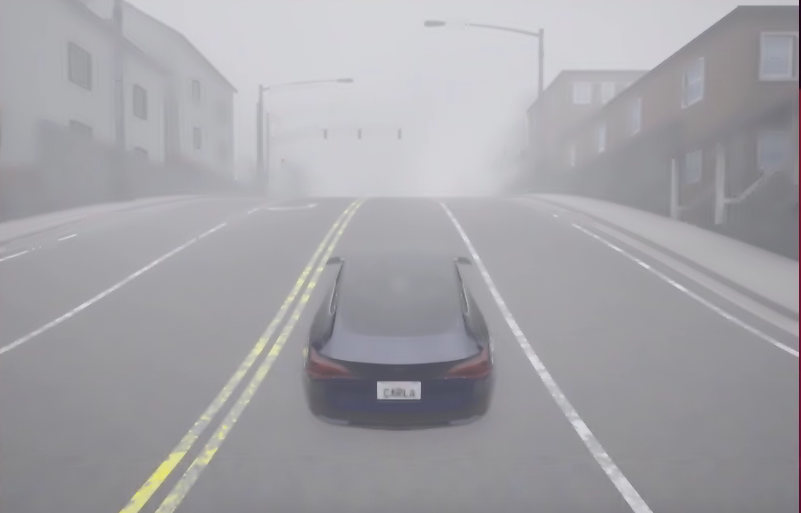}
		\label{fig:slope_pc}
	}
	\subfigure[]{
		\includegraphics[width=0.2\textwidth]{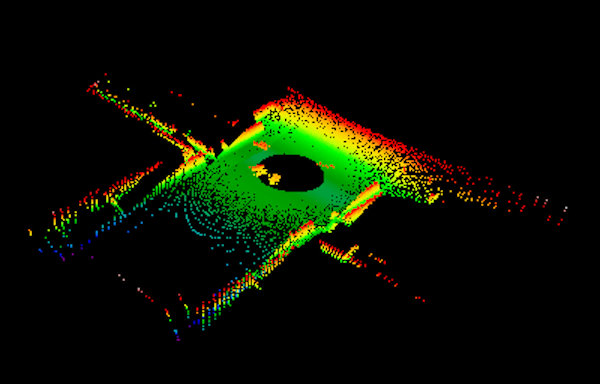}
		\label{fig:slope_camera}
	}
	
	\caption{Fog noise points in different scenarios.}
	\label{fig:point_cloud_noise}
\end{figure}

\autoref{fig:point_cloud_noise} illustrates the distribution pattern of fog noise points in different terrains. We set the fog visibility to 50 meters, and the color represents the height of the points. In \autoref{fig:junction_camera}, we place the ego vehicle in the middle of an open junction, and \autoref{fig:junction_pc} shows the point cloud captured by LiDAR. When the surroundings of the ego vehicle are open, fog noise points around the ego vehicle form a circle. As shown in \autoref{fig:bridge_camera} and \autoref{fig:bridge_pc}, if there are nearby buildings to the right of the ego vehicle, fog noise points will not occur on the right side because the signal reflected from nearby buildings is stronger than the signal reflected by fog droplets. At this point, fog noise points take on a semi-circular shape. When the ego vehicle is on a slope, as depicted in \autoref{fig:slope_camera} and \autoref{fig:slope_pc}, due to the upward tilt of the ego vehicle, a significant portion of the forward-emitted laser beams will go into the air, similar to when the ego vehicle is in an open area. In contrast, most of the backward-emitted lasers will hit the ground behind. In this case, fog noise points concentrate in the direct front of the ego vehicle.

\begin{figure}[]
	\centering
	\subfigure[Clear weather]{
		\includegraphics[width=0.2\textwidth]{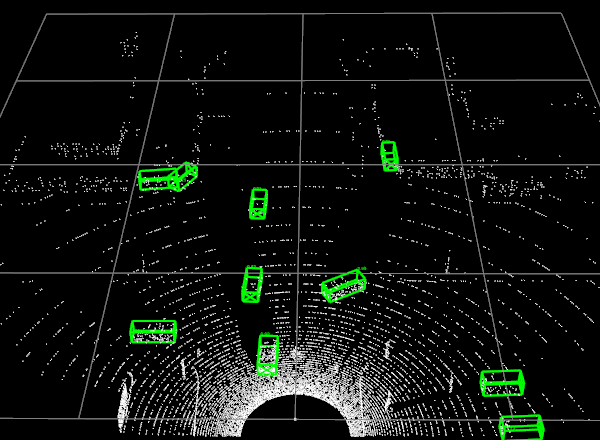}
		\label{fig:pvrcnn_clear}
	}
	\subfigure[200m visibility]{
		\includegraphics[width=0.2\textwidth]{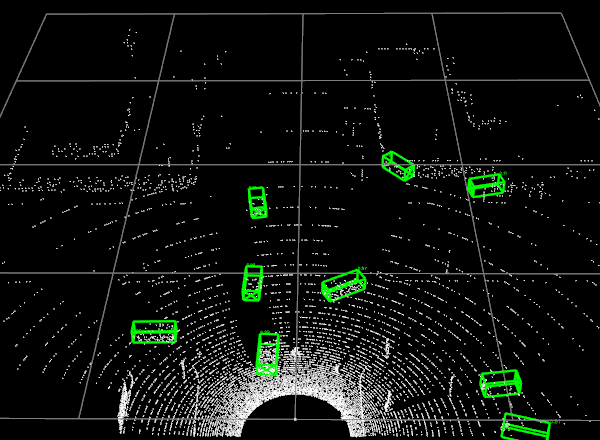}
		\label{fig:pvrcnn_200m}
	}
	
	\subfigure[50m visibility]{
		\includegraphics[width=0.2\textwidth]{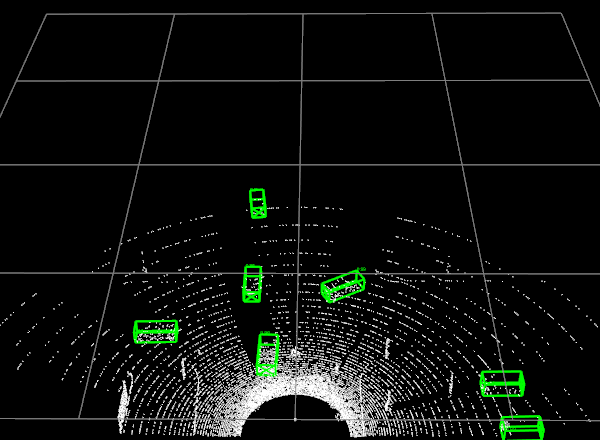}
		\label{fig:pvrcnn_50m}
	}
	\subfigure[20m visibility]{
		\includegraphics[width=0.2\textwidth]{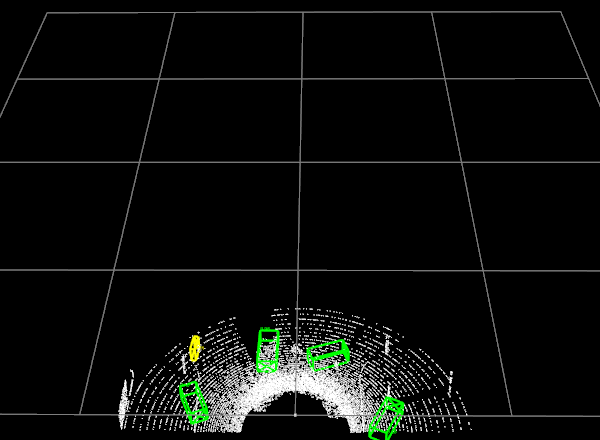}
		\label{fig:pvrcnn_20m}
	}
	
	\caption{Object detection performance of PV-RCNN under different visibility conditions.}
	\label{fig:3D_detection_foggy}
\end{figure}

To further illustrate the impact of fog, we placed some NPC vehicles around the ego vehicle. Representatively testing target detection algorithms with PV-RCNN~\cite{pv_rcnn}, as shown in \autoref{fig:3D_detection_foggy}. In \autoref{fig:pvrcnn_200m} and \autoref{fig:pvrcnn_50m}, as fog density increases, the detection range of the algorithm decreases. In \autoref{fig:pvrcnn_20m}, the detection range further decreases, and non-existent vehicles and pedestrians are detected near the ego vehicle. In conclusion, fog significantly affects LiDAR target detection results, causing a reduction in detection range and leading to false negatives and false positives in the detection results.

\subsection{Evaluation of ADSes in Predefined Foggy Scenarios}
\label{sec:leaderboard}

In this section, we evaluate the performance of ADSes in foggy weather using the CARLA Autonomous Driving Leaderboard~\cite{leaderboard}. The CARLA Leaderboard is a platform designed for assessing the performance of ADSes, comprising various predefined complex traffic scenarios based on the National Highway Traffic Safety Administration (NHTSA) pre-crash typology~\cite{NHTSA_precrash}. While CARLA Leaderboard scenarios encompass different weather conditions, the inability of CARLA itself to accurately simulate the impact of fog on LiDAR point clouds restricts its comprehensive testing of ADSes in foggy conditions.

We select eight representative scenarios from the Leaderboard for foggy weather testing, and the descriptions of these scenarios are provided at \url{https://idslab-autosec.github.io/}. Using both CARLA Leaderboard and \sys, we test InterFuser, TransFuser, and TF++, three autonomous driving algorithms employing MSF, in the same scenarios. Each scenario is tested 10 times, and pass rates are recorded.

As shown in \autoref{tab:leaderboard_result}, the performance drop of autonomous driving algorithms in foggy weather is more pronounced when using \sys compared to testing with CARLA Leaderboard. For instance, at a visibility of 50m, the pass rate for InterFuser is 76.3\% on CARLA Leaderboard, while it drops to 50\% on \sys.

In conclusion, the original CARLA LiDAR cannot accurately simulate the impact of fog on point clouds. Consequently, CARLA Leaderboard testing for ADSes in adverse weather is not comprehensive. In contrast, using \sys allows for a more realistic SOTIF testing of ADSes in adverse weather conditions, facilitating the discovery of related safety risks.

\begin{table}
	\caption{Pass rates of different autonomous driving algorithms on the CARLA Leaderboard in simulation testing.}
	\label{tab:leaderboard_result}
	\small
	\centering
	\begin{threeparttable}
		\begin{tabular}{ccccc}
			\toprule
		          \multirow{2}{*}{\textbf{Platform}} 
                    & \multirow{2}{*}{\textbf{Visibility}} 
                    & \multicolumn{3}{c}{\textbf{Autonomous Driving Algorithm}} \\
                        & & InterFuser & TransFuser & TF++\\ 
			\midrule
			\multirow{3}{*}{\shortstack{CARLA \\Leaderboard}}
                    & clear & 82.5\% & 77.5\% & 71.3\%\\
                    & 200m & 80.0\% & 72.5\% & 63.8\%\\
                    & 50m & 76.3\% & 68.8\% & 57.5\%\\
			\midrule
			\multirow{2}{*}{\sys}
                    & 200m & 68.8\% & 57.5\% & 48.8\%\\
                    & 50m & 50.0\% & 45.0\% & 41.3\%\\
			\bottomrule
		\end{tabular}
	\end{threeparttable}
\end{table}

\subsection{Corner Cases of Apollo OpenSource}

In \autoref{sec:leaderboard}, we conduct tests on scenarios manually defined on the CARLA Leaderboard. To enhance scenario diversity, we apply the corner case discovery algorithm outlined in \autoref{sec:scenario_generation} to Apollo OpenSource testing, revealing a series of corner cases.

\begin{figure}[htbp]
	\centering
	\subfigure[]{
		\includegraphics[width=0.13\textwidth]{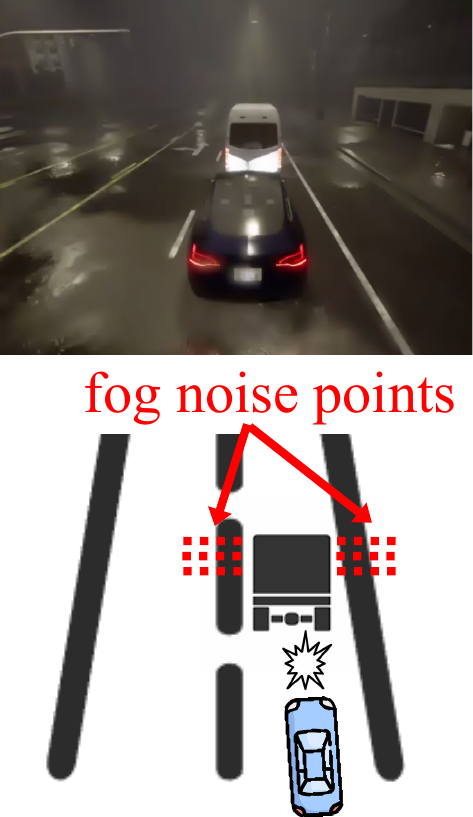}
		\label{fig:case_1}
	}
	\subfigure[]{
		\includegraphics[width=0.13\textwidth]{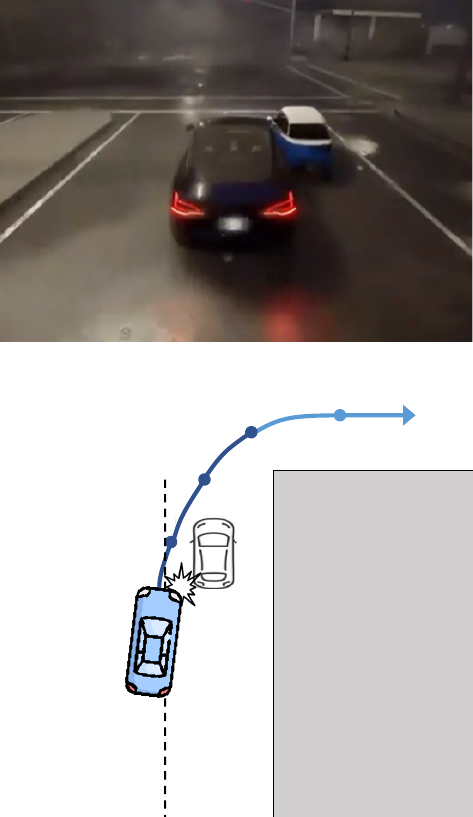}
		\label{fig:case_2}
	}
	\subfigure[]{
		\includegraphics[width=0.13\textwidth]{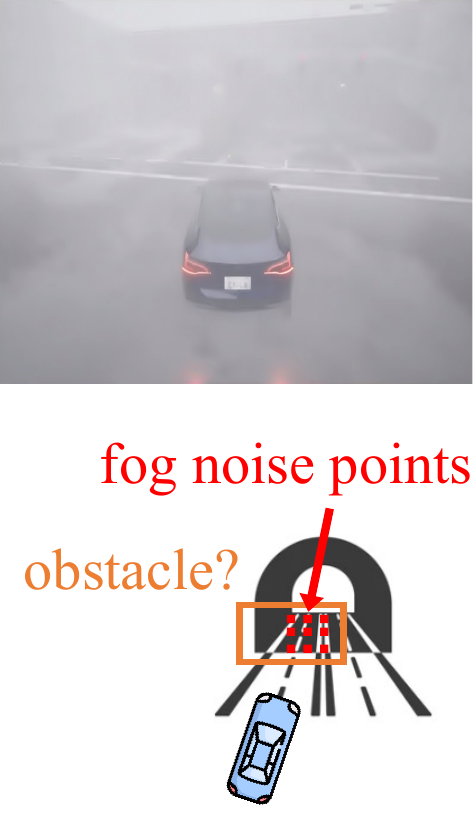}
		\label{fig:case_3}
	}
	\subfigure[]{
		\includegraphics[width=0.13\textwidth]{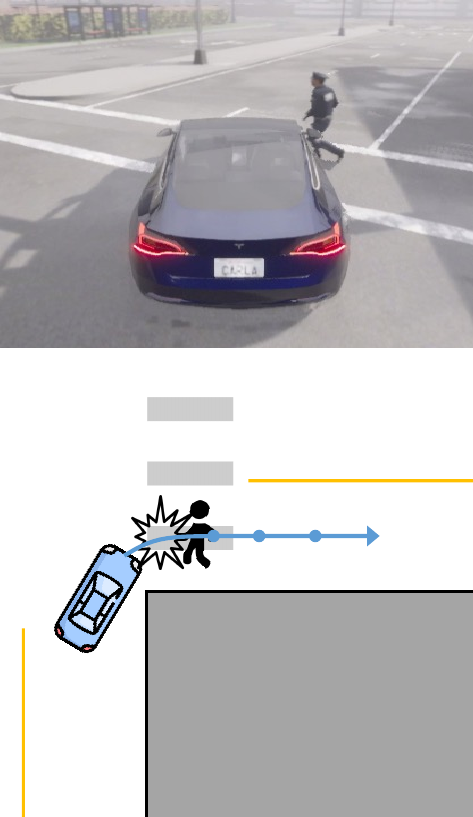}
		\label{fig:case_4}
	}
	\subfigure[]{
		\includegraphics[width=0.13\textwidth]{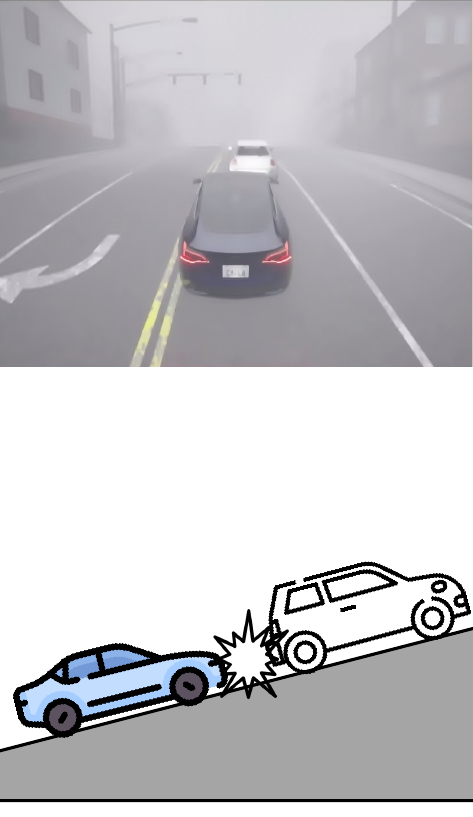}
		\label{fig:case_5}
	}
	\subfigure[]{
		\includegraphics[width=0.13\textwidth]{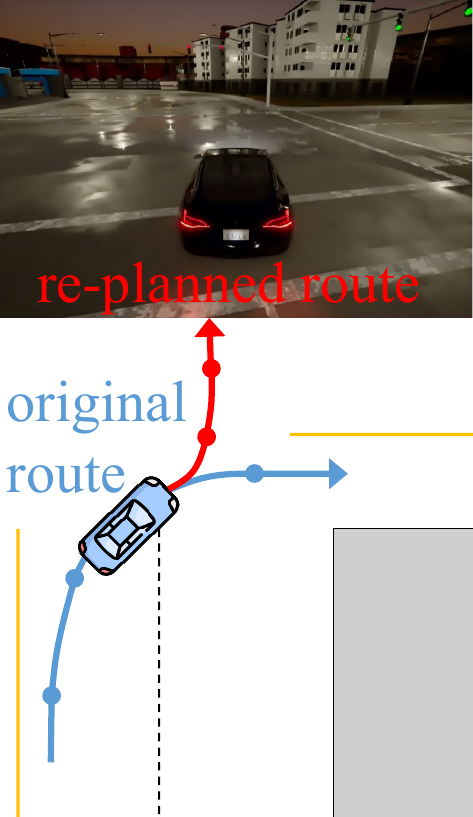}
		\label{fig:case_6}
	}
	\caption{Representative corner cases.}
	\label{fig:corner_case}
\end{figure}

\textbf{Case Study.} We analyze and discuss 6 representative corner cases, as shown in \autoref{fig:corner_case}. For a more in-depth analysis of additional corner cases and accompanying videos, please visit \url{https://idslab-autosec.github.io/}.

\begin{itemize}
	\item \autoref{fig:case_1} illustrates a corner case where the ego vehicle fails to detect a truck ahead in foggy weather. The fog interferes with the ego vehicle's LiDAR and camera, and the physical characteristics of the fog result in the fog noise points in the LiDAR point cloud having heights similar to those of the truck ahead. The fog noise points in the point cloud overlap significantly with the points from the truck, causing the ego vehicle to not detect the truck ahead, leading to a collision. If the weather is set to clear or the preceding vehicle model is changed to a shorter car, no collision occurs.
	\item In \autoref{fig:case_2}, there is a distant NPC vehicle ahead of the ego vehicle with a speed lower than that of the ego vehicle. The ego vehicle plans to change lanes to the right first and then make a right turn at the intersection. When the NPC vehicle is far away, the ego vehicle can detect it, so the ego vehicle plans to change lanes first and then follow the NPC vehicle in making a right turn. However, as the ego vehicle changes lanes, the distance to the NPC vehicle decreases, bringing the NPC vehicle into the range of fog noise points around the ego vehicle. At this point, the ego vehicle fails to detect the NPC vehicle, resulting in a collision.
	\item The corner case in \autoref{fig:case_3} is a false positive case in object detection. As the ego vehicle approaches a tunnel, the environment inside the tunnel is relatively open, and fog noise points mainly concentrate at the entrance of the tunnel. When the ego vehicle is close to the tunnel entrance, fog noise points connect with the stone wall on the left side of the tunnel entrance, being identified as a large obstacle. This misidentification causes the ego vehicle to come to a halt for an extended period at the tunnel entrance.
	\item In \autoref{fig:case_4}, the ego vehicle is making a right turn, and suddenly a pedestrian crosses the road. In foggy conditions, LiDAR's detection performance for small targets significantly decreases, leading to the ego vehicle not detecting the pedestrian and resulting in a collision. The same scenario doesn't result in a collision when the weather is clear.
	\item \autoref{fig:case_5} demonstrates a corner case on a slope. When the ego vehicle is ascending, the LiDAR angle is tilted, and fog noise points concentrate in front of the ego vehicle. This concentration of fog noise points leads to the ego vehicle failing to detect the vehicle ahead, resulting in a collision.
	\item The corner case in \autoref{fig:case_6} is primarily caused by the planning and localization modules. The ego vehicle plans to change lanes first and then make a right turn. When the ego vehicle reaches the intersection, its position is between the right-turn lane and the straight lane. However, at this moment, the ADS incorrectly judges that the ego vehicle is still in the straight lane, preventing it from making a right turn. The ego vehicle might end up going straight and then detouring back or may come to a stop at the intersection.
\end{itemize}

\begin{figure}[htbp]
	\centering
	\includegraphics[width=1.0\linewidth]{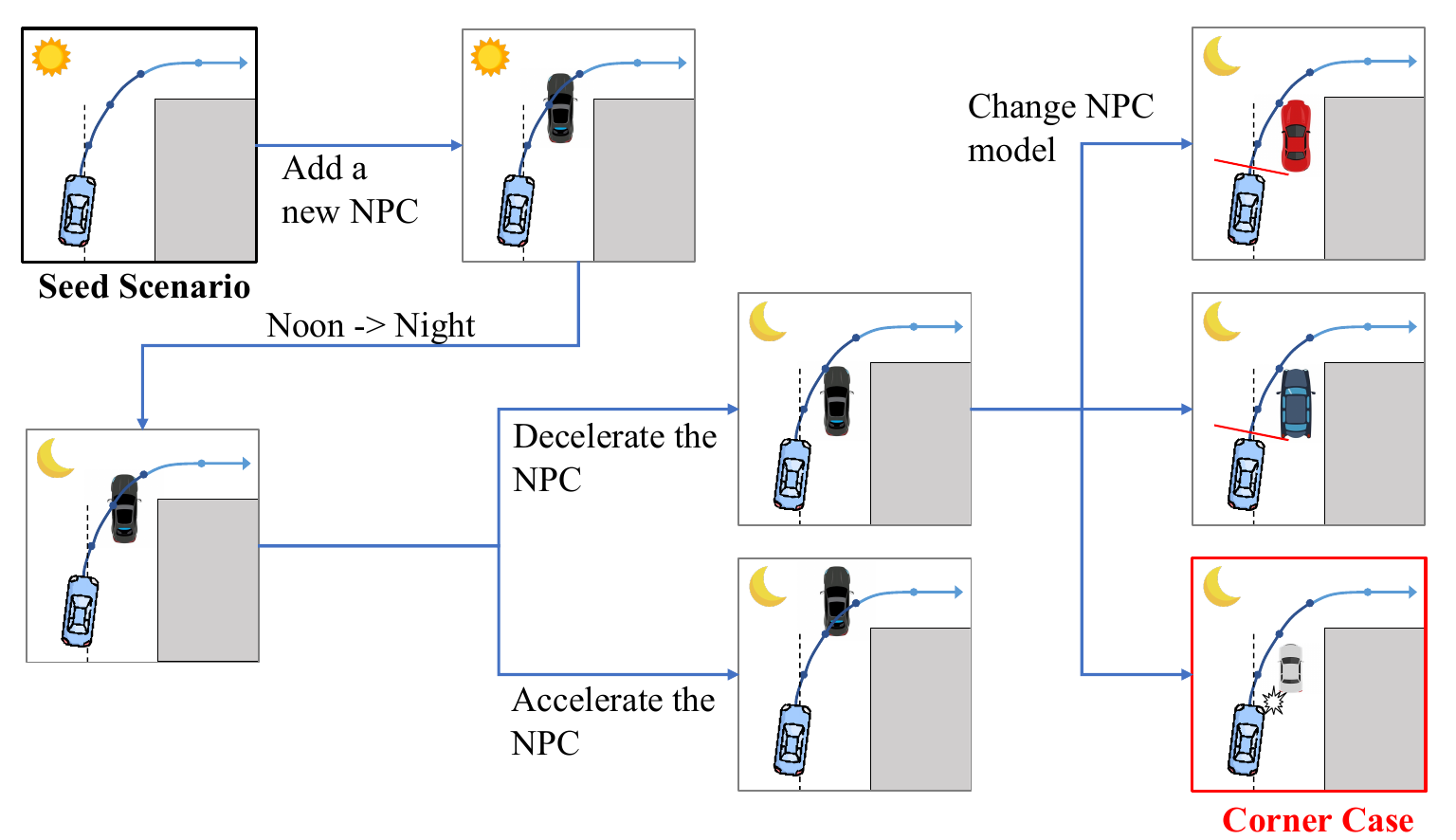}
	\caption{An example of scenario optimization.}
	\label{fig:optimization}
\end{figure}

\textbf{Example of Scenario Optimization.} In \autoref{sec:scenario_generation}, we design a simulated annealing algorithm to optimize scenarios, gradually modifying ordinary scenarios into corner cases. \autoref{fig:optimization} visually illustrates the process of modifying scenarios, with the weather set to foggy throughout the entire process. In the initial scenario, the ego vehicle makes a right turn with no vehicles or pedestrians ahead. Our algorithm initiates attempts to modify the scenario by introducing an NPC vehicle traveling straight ahead in front of the ego vehicle. Next, the sun altitude angle of the scenario is adjusted, changing the time from noon to night. This alteration in lighting conditions reduces the accuracy of camera recognition. In the next round, the algorithm attempts to accelerate and decelerate the NPC vehicle. When the NPC's speed decreases, the ego vehicle catches up, reducing the distance between them, resulting in a smaller objective function value. Therefore, the algorithm selects the scenario with reduced NPC speed. The algorithm then tries to modify the NPC's model, where in one instance, the NPC's vehicle type is changed from a regular-sized sedan to a smaller vehicle. This change causes the ego vehicle to inaccurately recognize the NPC vehicle, leading to a collision.

\textbf{Corner Cases of Autoware.} To investigate the transferability of the corner cases we identified, we replace the ADS with another open-source industrial-grade ADS, Autoware.Universe~\cite{autoware_universe}, and test it in a foggy environment. We observe that fog easily triggers false positives in Autoware's object detection. When visibility is below 100 meters, Autoware consistently misidentifies fog noise points as obstacles, leading to malfunctions. Consequently, we refrain from conducting further tests on Autoware.

\subsection{Effectiveness of Corner Case Discovery Algorithm}
Our objective function reflects the impact of fog on perception module and the collision probability during the ego vehicle's journey, enhancing the efficiency of finding corner cases. To validate this, we compare our objective function with a random objective function. Additionally, to demonstrate the efficiency of our algorithm in generating corner cases under foggy conditions, we compare our algorithm with DriveFuzz~\cite{drivefuzz}. We integrate Apollo into DriveFuzz and configure it with the same foggy environment.

\begin{figure}[htbp]
	\centering
	\includegraphics[width=1.0\linewidth]{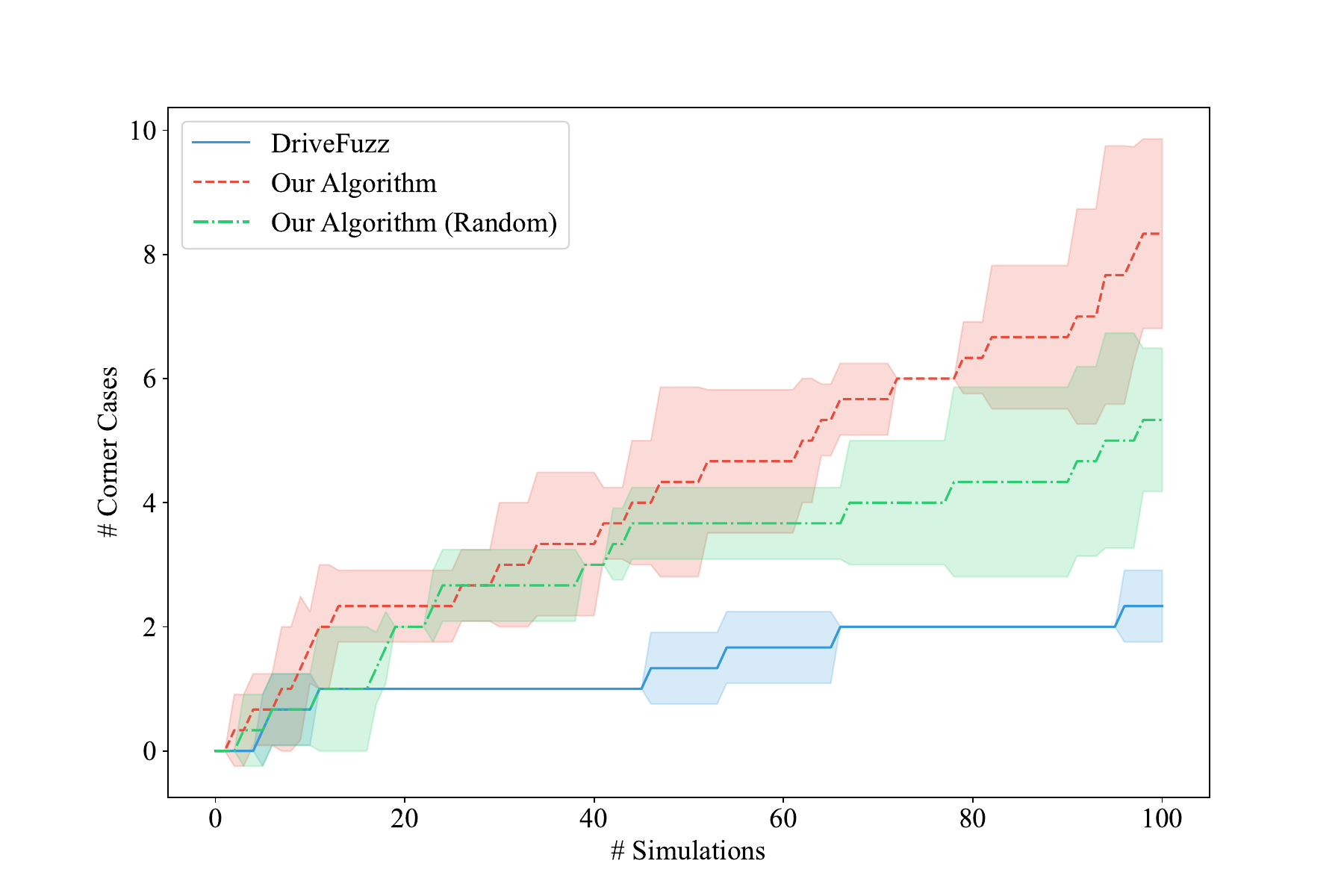}
	\caption{Number of Apollo's corner cases found in 100 simulations by different corner case discovery algorithms.}
	\label{fig:effectiveness}
\end{figure}

In \autoref{fig:effectiveness}, we run 100 simulations for each algorithm, repeat the process over 3 rounds, record the average and standard deviation of corner case counts, and manually validate each corner case, excluding false positives. The results highlight the superior efficiency of our algorithm in discovering corner cases compared to using a random objective function. Notably, our algorithm outperforms DriveFuzz in foggy conditions. This is attributed to our algorithm introducing safety-critical factors and enhancing NPC diversity by altering their models, whereas DriveFuzz relies on fixed NPC models. These findings underscore the importance of increasing NPC model diversity for more effective corner case discovery.

\section{Conclustion}
\label{sec:conclusion}
In this work, we introduce a novel sensor modeling approach base on first-principles, which has been effectively incorporated into the CARLA. This integration has unearthed a previously neglected set of corner cases pertaining to adverse weather conditions, and provided insights into their root causes. Guided by multiple empirical observations, we present a corner case discovery algorithm to augment the efficiency of scenario search. This approach not only offers guidance for initial scenarios and subsequent mutations but also notably reduces the dimensional space of the search, thereby boosting the efficiency in pinpointing corner cases. Our experimental findings confirm that, when compared with the state-of-the-art method under the same simulation parameters, our algorithm successfully discover approximately fourfold more corner cases within 100 simulation tests.
\section{Acknowledgmenent}

\begin{acks}

\end{acks}

\bibliographystyle{ACM-Reference-Format}
\bibliography{main}



\end{document}